\documentclass{bmcart}

\RequirePackage{natbib}
\usepackage{url}
\usepackage[utf8]{inputenc} 

\def\includegraphics{}

\startlocaldefs
\endlocaldefs

\begin{document}

\begin{frontmatter}

\begin{fmbox}
\dochead{Research}


\title{Aber-OWL: a framework for ontology-based data access
  in biology}




\author[
   addressref={aff1},                   
   email={roh25@aber.ac.uk}   
]{\inits{RH}\fnm{Robert} \snm{Hoehndorf}}
\author[
   addressref={aff1},
   email={lus11@aber.ac.uk}
]{\inits{LS}\fnm{Luke} \snm{Slater}}
\author[
   addressref={aff2},
   email={pns12@hermes.cam.ac.uk}
]{\inits{PNS}\fnm{Paul N} \snm{Schofield}}
\author[
   addressref={aff1},
   email={geg18@aber.ac.uk}
]{\inits{GVG}\fnm{Georgios V} \snm{Gkoutos}}


\address[id=aff1]{
  \orgname{Department of Computer Science, Aberystwyth University}, 
  \street{Llandinam Building},                     %
  \postcode{SY23 3DB}                                
  \city{Aberystwyth},                              
  \cny{UK}                                    
}
\address[id=aff2]{%
  \orgname{Department of Physiology, Development \& Neuroscience,
    University of Cambridge},
  \street{Downing Street},
  \postcode{CB2 3EG}
  \city{Cambridge},
  \cny{UK}
}



\end{fmbox}


\begin{abstractbox}

\begin{abstract} 
  Many ontologies have been developed in biology and these ontologies
  increasingly contain large volumes of formalized knowledge commonly
  expressed in the Web Ontology Language (OWL). Computational access
  to the knowledge contained within these ontologies relies on the use
  of automated reasoning.  We have developed the Aber-OWL
  infrastructure that provides reasoning services for
  bio-ontologies. Aber-OWL consists of an ontology repository, a set
  of web services and web interfaces that enable ontology-based
  semantic access to biological data and literature.  Aber-OWL is
  freely available at \url{http://aber-owl.net}.


\end{abstract}




\end{abstractbox}
%

\end{frontmatter}







\section*{Introduction}
A large number of ontologies have been developed for the annotation of
biological and biomedical data, commonly expressed in the Web Ontology
Language (OWL) \cite{Grau2008} or an OWL-compatible language such as
the OBO Flatfile Format \cite{Horrocks2007}. Access to the full extent
of knowledge contained in ontologies is facilitated by automated
reasoners that can compute the ontologies' underlying taxonomy and
answer queries over the ontology content.

While ontology repositories, such as BioPortal \cite{bioportal} and
the Ontology Lookup Service (OLS) \cite{ols}, provide web services and
interfaces to access ontologies, including their metadata such as
author names and licensing, the list of classes and asserted
structure, they do not enable computational access to the semantic
content of the ontologies and the inferences that can be drawn from
them.  Access to the semantic content of ontologies usually requires
further inferences to reveal the consequences of statements (axioms)
asserted in an ontology; these consequences may be automatically
derived using an automated reasoner.  To the best of our knowledge, no
reasoning infrastructure that supports semantically enabled access to
biological and biomedical ontologies currently exists.

Here, we present Aber-OWL, a reasoning infrastructure over ontologies
consisting of an ontology repository, web services that facilitate
semantic queries over ontologies specified by a user or contained in
Aber-OWL's repository, and a user interface. 
Such an infrastructure can not only enable access to knowledge
contained in ontologies, but crucially can also be used for semantic
queries over data annotated with ontologies, including the large
volumes of data that are increasingly becoming available through
public SPARQL endpoints \cite{Jupp2014}.  Allowing access to data
through an ontology is known as the ``ontology-based data access''
paradigm \cite{Bienvenu2013, Kontchakov2011}, and can exploit formal
information contained in ontologies to:
\begin{itemize}
\item identify possible inconsistencies and incoherent descriptions
  \cite{Hoehndorf2011incon},
\item enrich possibly incomplete data with background knowledge so as
  to obtain more complete answers to a query (e.g., if a data item
  referring to an organism has been characterized with findings of
  pulmonary stenosis, overriding aorta, ventricular septal defect, and
  right ventricular hypertrophy, and the ontology -- or the set of
  ontologies it imports -- contains enough information to allow, based
  on these four findings, the inference of a Tetralogy of Fallot
  condition, then the data item can be returned when querying for
  Tetralogy of Fallot even in the absence of it being explicitly
  declared in database) \cite{Hoehndorf2011phenome, Bienvenu2013},
\item enrich the data schema used to query data sources with
  additional information (e.g., by using a class in a query that is an
  inferred super-class of one or more classes that are used to
  annotate data items, but the class itself is never used to
  characterize data) \cite{Bienvenu2013}, and
\item provide a uniform view over multiple data sources with possibly
  heterogeneous, multi-modal data \cite{Bienvenu2013, Kontchakov2011}.
\end{itemize}

To demonstrate how Aber-OWL can be used for ontology-based access to
data, we provide a service that performs a semantic search over Pubmed
and Pubmed Central articles using the results of an Aber-OWL query,
and a service that performs SPARQL query extension so that the results
of Aber-OWL queries can be used to retrieve data accessible through
public SPARQL endpoints.  In Aber-OWL, following the ontology-based
data access paradigm \cite{Kontchakov2011, Bienvenu2013}, we specify
the features of the relevant information on the ontology- and
knowledge level \cite{Guarino1994}, and retrieve named classes in
ontologies satisfying these condition using an automated reasoner,
i.e., a software program that can identify whether a class in an
ontology satisfies certain conditions based on the axioms specified in
an ontology.
%
Subsequently, we embed the resulting information in database, Linked
Data or literature queries.

Aber-OWL can be accessed at \url{http://aber-owl.net}. The Aber-OWL
software is freely available at
\url{https://github.com/reality/SparqOWL} can be installed locally by
users who want to provide semantic access to their own ontologies and
support the use of their ontologies in semantic queries.

\section*{Materials and Methods}

\subsection*{Aber-OWL}


The Aber-OWL software can be configured with a list of URIs that
contain ontology documents (i.e., OWL files) and employs the OWL API
\cite{Horridge2007} to retrieve the ontologies that are to be included
in the repository. For each ontology document included in the
repository, the labels and definitions of all classes contained within
the ontology (as well as of all the ontologies it imports) are
identified based on OBO Foundry standards and recommendations: we use
the {\tt rdfs:label} annotation property to identify class labels for
each ontology (as well as of all the ontologies it imports), and we
employ the {\em definition} ({\tt
  http://purl.obo\-library.org/obo/IAO\_0000115}) annotation property,
defined in the Information Artifact Ontology, to identify the text
definitions of a class.
 
Labels of the classes occurring in each ontology, as well as of all
the ontologies it imports, are stored in a trie (prefix tree). The use
of a trie ensures that class labels can be searched efficiently, for
example when providing term completion recommendations.

Upon initiating the Aber-OWL web services, we classify each ontology
using the ELK reasoner \cite{Kazakov2011}, i.e., we identify the most
specific sub- and super-classes for each class contained in the
ontology using the axioms contained within it. The ELK reasoner
supports the OWL EL profile \cite{owlprofiles} and ignores ontology
axioms that do not fall within the OWL EL subset. The benefit of using
the OWL EL profile is the support for fast, polynomial-time reasoning,
and the OWL EL subset is a suitable dialect for a large number of
biomedical ontologies \cite{elvira}.  While we currently use ELK for
the Aber-OWL infrastructure, it is possible for a user to install an
Aber-OWL server that employs different OWL reasoners, such as HermiT
\cite{Motik2009a} or Pellet \cite{Sirin2004}, using the standard
reasoner interface of the OWL API.

Querying is performed by transforming a Manchester OWL Syntax
\cite{Horridge2006} query string into an OWL class expression using
the OWL API and then Aber-OWL's short-form provider is employed to
provide the mappings of the OWL class and the property URIs to the class
and property labels.  If this transformation fails (i.e., when the
query string provided is not a valid OWL class expression within the
ontology being queried), an empty set of results is returned. If the
transformation succeeds, the ELK reasoner is used to retrieve sub-,
super- or equivalent classes of the resulting OWL class
expression. The type of query (sub-class, super-class, or equivalent
class) is specified by the user and defaults to a sub-class
query. Queries in which the URL of the ontology document is not
specified are delegated to all ontologies in Aber-OWL's
repository. Consequently, results may be returned from different
ontologies. If a URL is specified as part of a query but the ontology
it corresponds to is not available within Aber-OWL's repository, an
attempt is made to retrieve the ontology from the URL, which is then
classified and then the query results over the classified ontology are
returned to the user. Should this process fail, an empty set of
results is returned. 

The results of an Aber-OWL query are provided in JSON format
\cite{Bray2014} and consist of an array of objects containing
information about the ontology classes satisfying the query: the URI
of the ontology document queried, the IRI of the ontology class, the
class label and the definition of the class. Detailed documentation of
the web services is available at the Aber-OWL web site.

We implemented a web server that can be used to access the Aber-OWL's
ontology repository and reasoning services. The web server features a
JQuery-based \cite{jquery} interface and uses AJAX \cite{Garrett2005} to retrieve
data from the Aber-OWL web services.

\subsection*{Aber-OWL: Pubmed}
Aber-OWL: Pubmed is built on top of the Aber-OWL reasoning
infrastructure.  It employes the Aber-OWL reasoning infrastructure
to resolve a semantic query formulated in Manchester OWL Syntax and
retrieve a set of named classes that satisfy the query. In particular,
depending on the type of query, all subclasses, superclasses or
equivalent classes that satisfy a class description in Manchester OWL
syntax within one or all ontologies in Aber-OWL's repository, or
within a user-specified ontology, are returned by Aber-OWL. The results
of the Aber-OWL query is a set of class descriptions, including the
class URI, the label and the definition of the class.  We use the
results to perform a Boolean textual search over a corpus of articles.

We use the Apache Lucene framework \cite{lucene} to create a fulltext
index of all titles and abstracts in MEDLINE/PubMed 2014
\cite{medline}, and all fulltext articles in Pubmed Central
\cite{pmc}. Before indexing, every text is processed using Lucene's
English language standard analyzer which tokenizes and normalises it
to lower case as well as applies a list of stop words.

For a user-specified query in Manchester OWL syntax, we construct a
Lucene query string from the set of class descriptions returned from
the Aber-OWL services. In particular, we concatenate each class label
using Lucene's {\tt OR} operator. As a result, the Lucene query will
match any article (title, abstract or fulltext) that contains a label
of a class satisfying the semantic query. It is also possible to
conjunctively perform multiple semantic queries by providing more than
one query in Manchester OWL syntax.

\subsection*{Aber-OWL: SPARQL}

Data in biology is commonly annotated to named classes in ontologies,
identified through a URI or another form of identifier that usually
directly maps to a URI. Pieces of data may refer to genes and
proteins, text passages, measurements and other observations, and can
be presented in multi-modal form as text, formal statements, images,
audio or video recordings. This information is increasingly being made
available as Linked Data through publicly available SPARQL endpoints
\cite{Jupp2014, Belleau2008}.

To semantically access ontology-annotated data contained in datasets
available through public SPARQL endpoints, we provide a service which
extends the SPARQL language with syntax which allows the user to
include Aber-OWL resultsets within the query. This comprises of a list
of class URIs returned by Aber-OWL, which can then be used to match
data in the SPARQL endpoint. SPARQL query expansion is implemented
using the PHP SPARQL library \cite{phpsparql} and is available both as a
web service and through a web interface that can be accessed through
Aber-OWL's main web site.


\section*{Results}

\subsection*{Aber-OWL}

The Aber-OWL framework can be used to retrieve all super-classes,
equivalent classes or sub-classes resulting from a Manchester OWL
Syntax query. The classes are retrieved either from a specific ontology in
Aber-OWL's ontology repository, from all ontologies in the repository,
or from a user-specified ontology that can be downloaded from a
specified URI.  In our installation of Aber-OWL at
\url{http://aber-owl.net}, the complete library of OBO ontologies
\cite{Smith2007} is imported as well as several user-requested
ontologies.

Using our web server, any ontology in Aber-OWL's ontology repository
can be queried and the results subsequently displayed.  Furthermore,
following execution of any Aber-OWL query, the web interface provides
the means to use the query in Aber-OWL: Pubmed so as to search and
retrieve relevant biomedical literature, or in Aber-OWL: SPARQL to
construct a query for data annotated to one of the resulting classes.

\subsection*{Ontology-based access to literature}

Aber-OWL: Pubmed enables ontology-based semantic access to biomedical
literature. It combines the information in biomedical ontologies with
automated reasoning to perform a literature query for all things that
can be inferred from a class description within one or more
ontologies. For example, a query for the class {\tt 'ventricular
  septal defect'} will return articles in which, among others, {\tt
  'tetralogy of fallot'} is mentioned due to {\tt 'tetralogy of
  fallot'} being inferred to be a subclass of {\tt 'ventricular septal
  defect'} in the Uberpheno \cite{sebastian2014} and Human Phenotype
\cite{Robinson2008} ontologies. Since Aber-OWL uses an automated
reasoner to identify subclasses, this information does not have to be
asserted in the ontology but rather is implied by the ontology's
axioms.

Aber-OWL: Pubmed can also perform more complex queries, such as for
articles containing mentions of subclasses of {\tt 'part\_of some
  'apoptotic process' and part\_of some regulation}, and articles
mentioning regulatory processes that are a part of apoptosis will be
returned. Such queries are only possible through the application of
automated reasoning over the knowledge contained in the biomedical
ontologies, and go beyond the state of the art in that they enable a
genuinely {\em semantic} way of accessing biomedical literature based
on the knowledge contained in the ontologies.

Finally, Aber-OWL: Pubmed can also be used to identify co-occurrences
of multiple Aber-OWL queries. For example, a conjunctive combination
of two sub-class queries, one for {\tt 'ventricular septal defect'}
and another for {\tt part\_of some heart}, will return articles that
contain references to both parts of the heart (such as the aorta) and
particular types of ventricular septal defects, e.g., muscular or
membranous defects, as well as complex phenotypes such as the
Tetralogy of Fallot.

Aber-OWL: Pubmed is accessible through a basic web interface at
\url{aber-owl.net/aber-owl/pubmed/} in which queries can be executed,
the articles satisfying the queries will be displayed, and matching
text passages in the title, abstract or fulltext will be
highlighted. Furthermore, Aber-OWL: Pubmed can be accessed through web
services and thereby can be embedded in web-based applications.

\subsection*{Ontology-based access to linked data}

Aber-OWL: SPARQL provides semantic access to Linked Data by expanding
SPARQL queries with the results returned by an Aber-OWL query. Query
expansion is performed based on SPARQL syntax extended by the
following construct:
\begin{verbatim}
OWL [querytype] [<Aber-OWL service URI>] [<ontology URI>] 
  { [OWL query] }
\end{verbatim}

For example, the query
\begin{verbatim}
OWL subclass <http://aber-owl.net/aber-owl/service/>
<http://purl.obolibrary.org/obo/go.owl> { part\_of some 'apoptotic
process' }
\end{verbatim}
will return a set of class URIs that satisfy the query {\tt part\_of
  some 'apoptotic process'} in the Gene Ontology (GO)
\cite{Ashburner2000}, and the results will be embedded in the SPARQL
query.  For this purpose, the {\tt OWL} statement is replaced by the
Aber-OWL: SPARQL service with a set of class URIs.  There are two main
forms in which the OWL statement can be embedded within a SPARQL
query. The first is the {\tt VALUES} form in which the results of the
OWL query are bound to a variable using the SPARQL 1.1 {\tt VALUES}
statement. For example,
\begin{verbatim}
VALUES ?ontid { 
  OWL subclass <http://aber-owl.net/aber-owl/service/> <> 
    { part_of some 'apoptotic process' }
}
\end{verbatim}
will bind the ontology URIs resulting from the OWL query ({\tt
  part\_of some 'apoptotic process'}) to the SPARQL variable {\tt
  ?ontid}.  The second form in which the OWL statement is useful is in
the form of a {\tt FILTER} statement. For example, the query
\begin{verbatim}
FILTER ( 
  ?ontid IN ( OWL subclass <http://aber-owl.net/aber-owl/service/> <> 
    { part_of some 'apoptotic process'} )
)
\end{verbatim}
will filter the results of a SPARQL query such that the values of {\tt
  ?ontid} must be in the result list of the OWL query.


As many SPARQL endpoints use different URIs to refer to classes in
ontologies, we have added the possibility to re-define prefixes for
the resulting ontology classes such that they match the IRI scheme
used by a particular SPARQL endpoint. When this feature is used, the
class IRIs resulting from an OWL query will be transformed into a
prefix form similar to the format used in the OBO Flatfile Format
\cite{Horrocks2007}, and the appropriate prefix definition will be
added to the SPARQL query if it has not been defined in the query
already. For example, the UniProt SPARQL endpoint
(\url{http://beta.sparql.uniprot.org}) uses the URI pattern {\tt
  http://purl.uniprot.org/go/<id>} to refer to Gene Ontology classes,
the EBI BioModels endpoint uses {\tt http://identifiers.org/go/<id>},
while the URI policy of the OBO Foundry \cite{obo-id-policy} specifies
that the URI pattern {\tt http://purl.obolibrary.org/obo/GO\_<id>}
should be used. The latter URI scheme is the one employed by Aber-OWL
since this is the authoritative URI provided in the ontology
document. Using the prefix format will transform the results of the
Aber-OWL query from URIs into strings of the type {\tt GO:<id>} and
the appropriate prefix to the SPARQL query (i.e., {\tt PREFIX GO:
  <http://purl.obolibrary.org/obo/GO\_>} will be added.  Changing this
prefix definition statement to {\tt PREFIX GO:
  <http://purl.uniprot.org/go/>} will effectively rewrite the URIs so
that they can be used in conjunction with the URI scheme employed by
the UniProt SPARQL endpoint.  Alternatively, the SPARQL query can
employ a dedicated mapping service, possibly in the form of a SPARQL
endpoint with access to {\tt sameAs} statements, to convert between
URI schemes used in different places.

\subsubsection*{Use case: Find all human proteins associated with a
  'part of apoptosis' in UniProt}
We can demonstrate the possibilities of using the Aber-OWL: SPARQL
query expansion service by retrieving all human proteins in UniProt
\cite{Uniprot2007} annotated to {\tt part\_of some 'apoptotic
  process'}. To achieve this goal, we use the SPARQL 1.1 {\tt VALUES}
statement to bind the results to a variable {\tt ?ontid}, and then we
can use this variable in the SPARQL query to retrieve all human
proteins with a Gene Ontology annotation in {\tt ?ontid}. The query is
shown in Figure \ref{fig:query-uniprot}.

As UniProt uses different URIs for GO classes than those returned by
Aber-OWL (which are based on the officially endorsed URIs by the OBO
Foundry and the Gene Ontology Consortium), the URIs have to be
rewritten for the query to succeed. In particular, in Aber-OWL:
SPARQL, an option must be activated to rewrite URIs into a ``prefix
form'' (i.e., URIs of the type {\tt
  http://purl.obolibrary.org/obo/GO\_0008150} would be rewritten to
{\tt GO:0008150}), and the SPARQL {\tt PREFIX} declaration will
redefine the prefix to match the URI scheme used in the UniProt SPARQL
endpoint. 

\begin{figure*}
\begin{verbatim}
PREFIX GO: <http://purl.uniprot.org/go/>
PREFIX taxon:<http://purl.uniprot.org/taxonomy/>
PREFIX up: <http://purl.uniprot.org/core/>
PREFIX skos: <http://www.w3.org/2004/02/skos/core#>

SELECT DISTINCT ?pname ?protein ?label ?ontid WHERE { 
  ##############################################
  # binds ?ontid to the results of the OWL query 
  VALUES ?ontid { 
    OWL subclass <http://aber-owl.net/aber-owl/service/> <>
        { part\_of some 'apoptotic process' }
  } . 
  ##############################################
  # ?ontid is now bound to the set of class IRIs of the OWL query 
        ?protein a up:Protein .
        ?protein up:organism taxon:9606 .
        ?protein up:mnemonic ?pname .
        ?protein up:classifiedWith ?ontid .
        ?ontid skos:prefLabel ?label .
}
\end{verbatim}
  \caption{\label{fig:query-uniprot}A query for all human proteins
    annotated to a part of apoptosis. The query is executed against
    the UniProt SPARQL endpoint at {\tt
      http://beta.sparql.uniprot.org}. To rewrite the URI scheme used
    by UniProt for GO classes to the URI scheme returned by Aber-OWL,
    Aber-OWL: SPARQL must be used with the prefix rewriting option set
    to {\tt true}.}
\end{figure*}

\subsubsection*{Use case: Search GWAS Central for genes and markers
  significantly involved in ventricular septal defects}
We can also utilize the Aber-OWL infrastructure for more powerful
queries that use inference over the ontology structure and utilize the
results in a SPARQL query. For example, we can use Aber-OWL: SPARQL to
query GWAS Central \cite{gwascentral} for markers that have been
identified in GWAS studies as significant for ventricular septal
defects. 

Using the Human Phenotype Ontology (HPO) \cite{Robinson2008} and the
definitions that have been developed for the HPO \cite{Mungall2010},
we can identify that a Tetralogy of Fallot is a particular type of
ventricular septal defect. In particular, according to the axioms
contained in the HPO, a Tetralogy of Fallot condition can be inferred
from the phenotypes {\em ventricular septal defect}, {\em overriding
  aorta}, {\em pulmonary valve stenosis} and {\em right ventricular
  hypertrophy}. Importantly, no explicit subclass relation between
these four key phenotypes and {\em Tetralogy of Fallot} is asserted in
the HPO. Therefore, reasoning is required to retrieve {\em Tetralogy
  of Fallot} as a subclass of either of these four, or a combination
of these four, phenotypes. Similarly, OWL reasoning over the ontology
axioms is required to retrieve data annotated to Tetralogy of Fallot
when querying for either of the four phenotypes. The queries can also
be made more precise by explicitly asking for a condition in which all
four of the Tetralogy of Fallot phenotypes must be satisfied:
subclasses of {\tt 'overriding aorta' and 'ventricular septal defect'
  and 'pulmonic stenosis' and 'right ventricular hypertrophy'} will
specifically retrieve the Tetralogy of Fallot condition, including
specific sub-types of Tetralogy of Fallot in the HPO.

\begin{figure*}
\begin{verbatim}
PREFIX rdf:<http://www.w3.org/1999/02/22-rdf-syntax-ns#>
PREFIX gc:<http://purl.org/gwas/schema#>
PREFIX xsd:<http://www.w3.org/2001/XMLSchema#>
PREFIX obo:<http://www.obofoundry.org/ro/ro.owl#>

SELECT ?gene ?ext_marker_id ?pvalue ?ontid
WHERE
{
    GRAPH ?g
    {
        ?marker gc:associated ?phenotype ;
        gc:locatedInGene ?gene ;
        gc:pvalue ?pvalue;
        obo:hasSynonym ?ext_marker_id.
        ?phenotype gc:hpoAnnotation ?ontid .
    }
    FILTER (xsd:float(?pvalue) <= 1e-10) .
    FILTER ( 
    ?ontid IN ( 
      OWL subclass <http://aber-owl.net/aber-owl/service/>
        <http://purl.obolibrary.org/obo/hp.owl>
          { 'ventricular septal defect' }
      )
    ) . 
}
\end{verbatim}
\caption{\label{fig:tof-sparql}A SPARQL query for markers
  significantly associated with {\em ventricular septal defect}. The
  query is executed against the GWAS Central SPARQL endpoint at {\tt
    http://fuseki.gwascentral.org/query.html}.}
\end{figure*}

\section*{Discussion}
\subsection*{Comparison to related work}

BioPortal \cite{bioportal}, the Ontology Lookup Service (OLS)
\cite{ols} and Ontobee \cite{Ontobee} are amongst the most widely used
ontology repositories in biology. These portals offer a user interface
for browsing ontologies and searching for classes based on the class
label (or synonym). They also provide web services that enable
programmatic access to the ontologies contained within them. However,
neither BioPortal, Ontobee nor OLS allow access to the knowledge that
can be derived from the ontologies in the repositories. Aber-OWL, on
the other hand, provides a reasoning infrastructure and services for
ontologies, without aiming at replacing ontology repositories and the
user experience they provide.  In the future, we intend to integrate
Aber-OWL more closely with other ontology repositories so that the
additional information and user-interface widgets provided by these
repositories can be combined with the reasoning infrastructure
provided by Aber-OWL.

Another related software is OntoQuery \cite{Tudose2013}, which is a
web-based query interface for ontologies that uses an OWL reasoner. It
can be used to provide an interface for a single ontology using an OWL
reasoner, but does not support use of multiple ontologies or access
through web interfaces.

The Logical Gene Ontology Annotations (GOAL) \cite{Jupp2012} outlines
an approach to access data annotated with ontologies through OWL
reasoning. For this purpose, GOAL constructs a custom knowledge base
integrating both the ontology and the annotations, and then uses an
OWL reasoner to answer queries over this combined knowledge
base. However, GOAL uses exactly one ontology, specifically built to
incorporate the data queried (mouse phenotypes) as a part of the OWL
ontology so that a reasoner can be used to query both, the ontology
and its annotations. Aber-OWL, on the other hand, is a general
framework and does not require changes to existing
ontologies. Instead, Aber-OWL distinguishes between reasoning on the
ontology level and retrieval of data annotated with ontologies.

Several tools and web servers utilize ontologies or structured
vocabularies for the retrieval of articles from Pubmed or
PubmedCentral. For example, GoPubmed \cite{Doms2005} classifies Pubmed
articles using the GO \cite{Ashburner2000} and the Medical Subjects
Heading thesaurus \cite{Nelson2004}.  However, GoPubmed uses only a
limited number of ontologies, and while GoPubmed uses the asserted
structure of the ontologies, it does not use the knowledge contained
within the ontologies' axioms. Aber-OWL: Pubmed, on the other hand,
can utilize the knowledge contained in any ontology to perform basic
searches in Pubmed abstracts and fulltext articles in Pubmed Central.

A main limitation of Aber-OWL: Pubmed lies with the absence of a
specialized entity recognition method to identify occurrences of
ontology class labels in text. In particular, for ontologies such as
the GO that use long and complex class names, specialized named entity
recognition approaches are required to identify mentions of the GO
terms in text \cite{Gaudan2008b, Blaschke2005}. Furthermore, Aber-OWL:
Pubmed currently uses only the {\tt rdfs:label} property of classes
and properties in ontologies to retrieve literature documents, but
ignores possible synonyms, alternative spellings or acronyms that may
be asserted for a class.  In the future, we will investigate the
possibility of adding more specialized named entity recognition
algorithms to Aber-OWL: Pubmed for specific ontologies.

Another limitation lies in Aber-OWL's interface. Aber-OWL: Pubmed's
web-based interface is not a complete text retrieval system but rather
demonstrates the possibility of using ontology-based queries for
retrieving text and can be used to aid in query construction. We
envision the main use of Aber-OWL: Pubmed in the form of its
web services that can be incorporated in more complete and more complex
text retrieval systems such as GoPubmed or even Pubmed itself.


The use of Aber-OWL: SPARQL differs in three key points from the use
of basic access to ontology-annotated data through SPARQL alone:
\begin{enumerate}
\item Aber-OWL: SPARQL provides access to the semantic content of
  ontologies even when the ontologies are not available through the
  SPARQL endpoint that contains the ontology-annotated data.
\item Aber-OWL: SPARQL provides access to the inferred ontology
  structure instead of the asserted structure, even when no OWL
  entailment regime is activated in a SPARQL endpoint.
\item Aber-OWL: SPARQL enables complex queries formulated in
  Manchester OWL syntax, and can perform these queries even when no
  OWL entailment regime is activated in a SPARQL endpoint.
\end{enumerate}

In particular, (1) the ontologies used for annotation are not commonly
accessible through the same SPARQL endpoint as the actual annotated
data. If the SPARQL endpoint supports query federation (using the
SPARQL {\tt SERVICE} block), this problem can usually be resolved if
the ontology is available at some place (such as BioPortal) through
another SPARQL endpoint. However, in some application settings, a
query expansion service may be more efficient than query federation.
More importantly, however, (2) Aber-OWL: SPARQL provides access to the
structure of an ontology as it is inferred by an OWL reasoner. To
achieve a similar outcome using plain SPARQL, the SPARQL endpoint
containing the ontology must have an OWL entailment regime
\cite{sparqlentailment} activated; otherwise, only the asserted
structure of an ontology is available for queries. We know of no
SPARQL endpoint in the biomedical domain currently holding ontologies
and simultaneously using an OWL entailment regime; in particular,
neither BioPortal nor Ontobee or the OLS currently make use of any
kind of OWL entailment. While the first two points can in principle be
addressed by applying Semantic Web technologies, queries would still
have to be formulated in SPARQL syntax. (3) Aber-OWL: SPARQL uses the
Manchester OWL syntax to formulate queries, and Manchester OWL syntax
is widely used by ontology developers and users as it is closer to a
human-readable sentence and therefore easier to access than other ways
of expressing OWL.

\subsection*{The need for improved interoperability between biomedical ontologies}

The full benefit of a reasoning infrastructure over multiple
ontologies can be realized when these ontologies are
``interoperable''.  While interoperability between biomedical
ontologies has been extensively discussed \cite{Smith2005, Smith2007,
  Koehler2011, Hoehndorf2011incon}, we can nevertheless identify
several shortcomings through the use of Aber-OWL.  Firstly, ontology
class names and relation names are not standardized.  For example, the
current library of ontologies included in Aber-OWL uses several
different names (and URIs) for the {\tt part-of} relation, including
{\tt part\_of}, {\tt part-of}, {\tt 'part of'} and {\tt PartOf}. While
each relation is usually consistently applied within a single
ontology, the use of different URIs and labels for the same relation
leads to difficulties when utilizing more than one ontology. The
non-standardized use of relation names is particularly surprising as
the OBO Relation Ontology \cite{Smith2005} aimed to achieve the goal
of using standard relations and common relation names almost 10 years
ago. One possible explanation for the observed heterogeneity is that
the lack of tools and an infrastructure that could efficiently utilize
the information in one or more ontology has made it less of a priority
for ontology developers to focus on these aspects of interoperability.

Furthermore, using the Aber-OWL infrastructure, potential problems in
ontologies can be identified. For example, we could identify, and
subsequently correct, three unsatisfiable classes in the Neuro
Behavior Ontology \cite{Hoehndorf2013nbo} resulting from changes in
the ontologies it imports. These problems are not easily detectable;
moreover, they require the use of reasoning over more than one
ontology, as well as frequent re-classifications. These tasks are
vital for the effects that a change in one ontology has on other
ontologies to be detected.

\subsection*{The ontology-based data access paradigm}

With the Aber-OWL services, we propose to separate the processing of
knowledge in ontologies and the retrieval of data annotated with these
ontologies. Aber-OWL provides a reasoning infrastructure that can be
queried either through its web interface or its web services, and a
set of classes that satisfy a specified condition is returned. These
sets of classes can then be used to retrieve data annotated with them,
text that contains their label, or from a corpus of text or a formal
data resource that references them. As such, Aber-OWL provides a
framework for automatically accessing information that is annotated
with ontologies or contains terms used to label classes in
ontologies. When using Aber-OWL, access to the information in
ontologies is not merely based on class names or identifiers but
rather on the knowledge the ontologies contain and the inferences that
can be drawn from it. This also enables the use of knowledge- and
ontology-based access to data \cite{Bienvenu2013, Kontchakov2011}:
data of interest is specified on the knowledge- or ontology-level
\cite{Guarino1994}, and all possible classes that satisfy such a
specification are inferred using an automated reasoner. The results of
this inference process are then used to actually retrieve the data
without the need to apply further inference.




\section*{Acknowledgements}

\bibliographystyle{bmc-mathphys}

\begin{thebibliography}{44}
\ifx \bisbn   \undefined \def \bisbn  #1{ISBN #1}\fi
\ifx \binits  \undefined \def \binits#1{#1}\fi
\ifx \bauthor  \undefined \def \bauthor#1{#1}\fi
\ifx \batitle  \undefined \def \batitle#1{#1}\fi
\ifx \bjtitle  \undefined \def \bjtitle#1{#1}\fi
\ifx \bvolume  \undefined \def \bvolume#1{\textbf{#1}}\fi
\ifx \byear  \undefined \def \byear#1{#1}\fi
\ifx \bissue  \undefined \def \bissue#1{#1}\fi
\ifx \bfpage  \undefined \def \bfpage#1{#1}\fi
\ifx \blpage  \undefined \def \blpage #1{#1}\fi
\ifx \burl  \undefined \def \burl#1{\textsf{#1}}\fi
\ifx \doiurl  \undefined \def \doiurl#1{\textsf{#1}}\fi
\ifx \betal  \undefined \def \betal{\textit{et al.}}\fi
\ifx \binstitute  \undefined \def \binstitute#1{#1}\fi
\ifx \binstitutionaled  \undefined \def \binstitutionaled#1{#1}\fi
\ifx \bctitle  \undefined \def \bctitle#1{#1}\fi
\ifx \beditor  \undefined \def \beditor#1{#1}\fi
\ifx \bpublisher  \undefined \def \bpublisher#1{#1}\fi
\ifx \bbtitle  \undefined \def \bbtitle#1{#1}\fi
\ifx \bedition  \undefined \def \bedition#1{#1}\fi
\ifx \bseriesno  \undefined \def \bseriesno#1{#1}\fi
\ifx \blocation  \undefined \def \blocation#1{#1}\fi
\ifx \bsertitle  \undefined \def \bsertitle#1{#1}\fi
\ifx \bsnm \undefined \def \bsnm#1{#1}\fi
\ifx \bsuffix \undefined \def \bsuffix#1{#1}\fi
\ifx \bparticle \undefined \def \bparticle#1{#1}\fi
\ifx \barticle \undefined \def \barticle#1{#1}\fi
\ifx \bconfdate \undefined \def \bconfdate #1{#1}\fi
\ifx \botherref \undefined \def \botherref #1{#1}\fi
\ifx \url \undefined \def \url#1{\textsf{#1}}\fi
\ifx \bchapter \undefined \def \bchapter#1{#1}\fi
\ifx \bbook \undefined \def \bbook#1{#1}\fi
\ifx \bcomment \undefined \def \bcomment#1{#1}\fi
\ifx \oauthor \undefined \def \oauthor#1{#1}\fi
\ifx \citeauthoryear \undefined \def \citeauthoryear#1{#1}\fi
\ifx \endbibitem  \undefined \def \endbibitem {}\fi
\ifx \bconflocation  \undefined \def \bconflocation#1{#1}\fi
\ifx \arxivurl  \undefined \def \arxivurl#1{\textsf{#1}}\fi
\csname PreBibitemsHook\endcsname

\bibitem{Grau2008}
\begin{barticle}
\bauthor{\bsnm{Grau}, \binits{B.}},
\bauthor{\bsnm{Horrocks}, \binits{I.}},
\bauthor{\bsnm{Motik}, \binits{B.}},
\bauthor{\bsnm{Parsia}, \binits{B.}},
\bauthor{\bsnm{Patelschneider}, \binits{P.}},
\bauthor{\bsnm{Sattler}, \binits{U.}}:
\batitle{{OWL} 2: The next step for {OWL}}.
\bjtitle{Web Semantics: Science, Services and Agents on the World Wide Web}
\bvolume{6}(\bissue{4}),
\bfpage{309}--\blpage{322}
(\byear{2008})
\end{barticle}
\endbibitem

\bibitem{Horrocks2007}
\begin{botherref}
\oauthor{\bsnm{Horrocks}, \binits{I.}}:
{OBO} flat file format syntax and semantics and mapping to {OWL} {W}eb
  {O}ntology {L}anguage.
Technical report,
University of Manchester
(March 2007).
\url{http://www.cs.man.ac.uk/~horrocks/obo/}
\end{botherref}
\endbibitem

\bibitem{bioportal}
\begin{barticle}
\bauthor{\bsnm{Noy}, \binits{N.F.}},
\bauthor{\bsnm{Shah}, \binits{N.H.}},
\bauthor{\bsnm{Whetzel}, \binits{P.L.}},
\bauthor{\bsnm{Dai}, \binits{B.}},
\bauthor{\bsnm{Dorf}, \binits{M.}},
\bauthor{\bsnm{Griffith}, \binits{N.}},
\bauthor{\bsnm{Jonquet}, \binits{C.}},
\bauthor{\bsnm{Rubin}, \binits{D.L.}},
\bauthor{\bsnm{Storey}, \binits{M.-A.A.}},
\bauthor{\bsnm{Chute}, \binits{C.G.}},
\bauthor{\bsnm{Musen}, \binits{M.A.}}:
\batitle{Bioportal: ontologies and integrated data resources at the click of a
  mouse.}
\bjtitle{Nucleic acids research}
\bvolume{37}(\bissue{Web Server issue}),
\bfpage{170}--\blpage{173}
(\byear{2009}).
doi:\url{10.1093/nar/gkp440}
\end{barticle}
\endbibitem

\bibitem{ols}
\begin{barticle}
\bauthor{\bsnm{Cote}, \binits{R.}},
\bauthor{\bsnm{Jones}, \binits{P.}},
\bauthor{\bsnm{Apweiler}, \binits{R.}},
\bauthor{\bsnm{Hermjakob}, \binits{H.}}:
\batitle{The ontology lookup service, a lightweight cross-platform tool for
  controlled vocabulary queries}.
\bjtitle{BMC Bioinformatics}
\bvolume{7}(\bissue{1}),
\bfpage{97}
(\byear{2006}).
doi:\url{10.1186/1471-2105-7-97}
\end{barticle}
\endbibitem

\bibitem{Jupp2014}
\begin{barticle}
\bauthor{\bsnm{Jupp}, \binits{S.}},
\bauthor{\bsnm{Malone}, \binits{J.}},
\bauthor{\bsnm{Bolleman}, \binits{J.}},
\bauthor{\bsnm{Brandizi}, \binits{M.}},
\bauthor{\bsnm{Davies}, \binits{M.}},
\bauthor{\bsnm{Garcia}, \binits{L.}},
\bauthor{\bsnm{Gaulton}, \binits{A.}},
\bauthor{\bsnm{Gehant}, \binits{S.}},
\bauthor{\bsnm{Laibe}, \binits{C.}},
\bauthor{\bsnm{Redaschi}, \binits{N.}},
\bauthor{\bsnm{Wimalaratne}, \binits{S.M.}},
\bauthor{\bsnm{Martin}, \binits{M.}},
\bauthor{\bsnm{Le~Novère}, \binits{N.}},
\bauthor{\bsnm{Parkinson}, \binits{H.}},
\bauthor{\bsnm{Birney}, \binits{E.}},
\bauthor{\bsnm{Jenkinson}, \binits{A.M.}}:
\batitle{The {EBI} {RDF} platform: linked open data for the life sciences}.
\bjtitle{Bioinformatics}
\bvolume{30}(\bissue{9}),
\bfpage{1338}--\blpage{1339}
(\byear{2014}).
doi:\url{10.1093/bioinformatics/btt765}
\end{barticle}
\endbibitem

\bibitem{Bienvenu2013}
\begin{bchapter}
\bauthor{\bsnm{Bienvenu}, \binits{M.}},
\bauthor{\bparticle{ten} \bsnm{Cate}, \binits{B.}},
\bauthor{\bsnm{Lutz}, \binits{C.}},
\bauthor{\bsnm{Wolter}, \binits{F.}}:
\bctitle{Ontology-based data access: a study through disjunctive datalog, csp,
  and mmsnp}.
In: \bbtitle{PODS},
pp. \bfpage{213}--\blpage{224}
(\byear{2013})
\end{bchapter}
\endbibitem

\bibitem{Kontchakov2011}
\begin{bchapter}
\bauthor{\bsnm{Kontchakov}, \binits{R.}},
\bauthor{\bsnm{Lutz}, \binits{C.}},
\bauthor{\bsnm{Toman}, \binits{D.}},
\bauthor{\bsnm{Wolter}, \binits{F.}},
\bauthor{\bsnm{Zakharyaschev}, \binits{M.}}:
\bctitle{The combined approach to ontology-based data access}.
In: \bbtitle{IJCAI},
pp. \bfpage{2656}--\blpage{2661}
(\byear{2011})
\end{bchapter}
\endbibitem

\bibitem{Hoehndorf2011incon}
\begin{barticle}
\bauthor{\bsnm{Hoehndorf}, \binits{R.}},
\bauthor{\bsnm{Dumontier}, \binits{M.}},
\bauthor{\bsnm{Oellrich}, \binits{A.}},
\bauthor{\bsnm{Rebholz-Schuhmann}, \binits{D.}},
\bauthor{\bsnm{Schofield}, \binits{P.N.}},
\bauthor{\bsnm{Gkoutos}, \binits{G.V.}}:
\batitle{Interoperability between biomedical ontologies through relation
  expansion, upper-level ontologies and automatic reasoning}.
\bjtitle{PLOS ONE}
\bvolume{6}(\bissue{7}),
\bfpage{22006}
(\byear{2011})
\end{barticle}
\endbibitem

\bibitem{Hoehndorf2011phenome}
\begin{barticle}
\bauthor{\bsnm{Hoehndorf}, \binits{R.}},
\bauthor{\bsnm{Schofield}, \binits{P.N.}},
\bauthor{\bsnm{Gkoutos}, \binits{G.V.}}:
\batitle{Phenomenet: a whole-phenome approach to disease gene discovery}.
\bjtitle{Nucleic Acids Research}
\bvolume{39}(\bissue{18}),
\bfpage{119}
(\byear{2011})
\end{barticle}
\endbibitem

\bibitem{Guarino1994}
\begin{bchapter}
\bauthor{\bsnm{Guarino}, \binits{N.}}:
\bctitle{The ontological level}.
In: \beditor{\bsnm{Casati}, \binits{R.}},
\beditor{\bsnm{Smith}, \binits{B.}},
\beditor{\bsnm{White}, \binits{G.}} (eds.)
\bbtitle{Philosophy and the Cognitive Sciences},
pp. \bfpage{443}--\blpage{456}.
\bpublisher{H\"{o}lder-Pichler-Tempsky},
\blocation{Vienna}
(\byear{1994})
\end{bchapter}
\endbibitem

\bibitem{Horridge2007}
\begin{bchapter}
\bauthor{\bsnm{Horridge}, \binits{M.}},
\bauthor{\bsnm{Bechhofer}, \binits{S.}},
\bauthor{\bsnm{Noppens}, \binits{O.}}:
\bctitle{Igniting the {OWL} 1.1 touch paper: The {OWL} {API}}.
In: \bbtitle{Proceedings of OWLED 2007: Third International Workshop on OWL
  Experiences and Directions}
(\byear{2007})
\end{bchapter}
\endbibitem

\bibitem{Kazakov2011}
\begin{bchapter}
\bauthor{\bsnm{Kazakov}, \binits{Y.}},
\bauthor{\bsnm{Kr{\"o}tzsch}, \binits{M.}},
\bauthor{\bsnm{Siman\v{c}\'{i}k}, \binits{F.}}:
\bctitle{Unchain my $\mathcal{EL}$ reasoner}.
In: \bbtitle{Proceedings of the 23rd International Workshop on Description
  Logics (DL'10)}.
\bsertitle{CEUR Workshop Proceedings}.
\bpublisher{CEUR-WS.org}, \blocation{???}
(\byear{2011})
\end{bchapter}
\endbibitem

\bibitem{owlprofiles}
\begin{botherref}
\oauthor{\bsnm{Motik}, \binits{B.}},
\oauthor{\bsnm{Grau}, \binits{B.C.}},
\oauthor{\bsnm{Horrocks}, \binits{I.}},
\oauthor{\bsnm{Wu}, \binits{Z.}},
\oauthor{\bsnm{Fokoue}, \binits{A.}},
\oauthor{\bsnm{Lutz}, \binits{C.}}:
Owl 2 web ontology language: Profiles.
Recommendation,
World Wide Web Consortium (W3C)
(2009)
\end{botherref}
\endbibitem

\bibitem{elvira}
\begin{barticle}
\bauthor{\bsnm{Hoehndorf}, \binits{R.}},
\bauthor{\bsnm{Dumontier}, \binits{M.}},
\bauthor{\bsnm{Oellrich}, \binits{A.}},
\bauthor{\bsnm{Wimalaratne}, \binits{S.}},
\bauthor{\bsnm{Rebholz-Schuhmann}, \binits{D.}},
\bauthor{\bsnm{Schofield}, \binits{P.}},
\bauthor{\bsnm{Gkoutos}, \binits{G.V.}}:
\batitle{A common layer of interoperability for biomedical ontologies based on
  {OWL} {EL}}.
\bjtitle{Bioinformatics}
\bvolume{27}(\bissue{7}),
\bfpage{1001}--\blpage{1008}
(\byear{2011})
\end{barticle}
\endbibitem

\bibitem{Motik2009a}
\begin{barticle}
\bauthor{\bsnm{Motik}, \binits{B.}},
\bauthor{\bsnm{Shearer}, \binits{R.}},
\bauthor{\bsnm{Horrocks}, \binits{I.}}:
\batitle{{Hypertableau Reasoning for Description Logics}}.
\bjtitle{Journal of Artificial Intelligence Research}
\bvolume{36},
\bfpage{165}--\blpage{228}
(\byear{2009})
\end{barticle}
\endbibitem

\bibitem{Sirin2004}
\begin{bchapter}
\bauthor{\bsnm{Sirin}, \binits{E.}},
\bauthor{\bsnm{Parsia}, \binits{B.}}:
\bctitle{Pellet: An {OWL} {DL} reasoner}.
In: \beditor{\bsnm{Haarslev}, \binits{V.}},
\beditor{\bsnm{M{\"{o}}ller}, \binits{R.}} (eds.)
\bbtitle{Proceedings of the 2004 International Workshop on Description Logics,
  DL2004, Whistler, British Columbia, Canada, Jun 6-8}.
\bsertitle{CEUR Workshop Proceedings},
vol. \bseriesno{104}.
\bpublisher{CEUR-WS.org},
\blocation{Aachen, Germany}
(\byear{2004})
\end{bchapter}
\endbibitem

\bibitem{Horridge2006}
\begin{botherref}
\oauthor{\bsnm{Horridge}, \binits{M.}},
\oauthor{\bsnm{Drummond}, \binits{N.}},
\oauthor{\bsnm{Goodwin}, \binits{J.}},
\oauthor{\bsnm{Rector}, \binits{A.}},
\oauthor{\bsnm{Stevens}, \binits{R.}},
\oauthor{\bsnm{Wang}, \binits{H.H.}}:
{The Manchester OWL Syntax}.
Proc. of the 2006 OWL Experiences and Directions Workshop (OWL-ED2006)
(2006)
\end{botherref}
\endbibitem

\bibitem{Bray2014}
\begin{botherref}
\oauthor{\bsnm{Bray}, \binits{T.}}:
{The JavaScript Object Notation (JSON) Data Interchange Format}.
IETF
(2014).
\url{http://www.ietf.org/rfc/rfc7159.txt}
\end{botherref}
\endbibitem

\bibitem{jquery}
\begin{botherref}
\oauthor{\bsnm{{The jQuery Project}}}:
{jQuery}: The write less, do more, JavaScript library.
\url{http://jquery.com}
\end{botherref}
\endbibitem

\bibitem{Garrett2005}
\begin{botherref}
\oauthor{\bsnm{Garrett}, \binits{J.J.}}:
Ajax: A New Approach to Web Applications.
\url{http://www.adaptivepath.com/ideas/essays/archives/000385.php}
(2005)
\end{botherref}
\endbibitem

\bibitem{lucene}
\begin{botherref}
\oauthor{\bsnm{{The Apache Software Foundation}}}:
Apache Lucene.
\url{http://lucene.apache.org}
\end{botherref}
\endbibitem

\bibitem{medline}
\begin{botherref}
\oauthor{\bsnm{{U.S. National Library of Medicine}}}:
2014 MEDLINE/PubMed Baseline Distribution.
\url{http://www.nlm.nih.gov/bsd/licensee/2014_stats/baseline_doc.html}
\end{botherref}
\endbibitem

\bibitem{pmc}
\begin{botherref}
\oauthor{\bsnm{{U.S. National Library of Medicine}}}:
PubMed Central.
\url{http://www.ncbi.nlm.nih.gov/pmc/}
\end{botherref}
\endbibitem

\bibitem{Belleau2008}
\begin{barticle}
\bauthor{\bsnm{Belleau}, \binits{F.}},
\bauthor{\bsnm{Nolin}, \binits{M.}},
\bauthor{\bsnm{Tourigny}, \binits{N.}},
\bauthor{\bsnm{Rigault}, \binits{P.}},
\bauthor{\bsnm{Morissette}, \binits{J.}}:
\batitle{{Bio2RDF}: Towards a mashup to build bioinformatics knowledge
  systems}.
\bjtitle{Journal of Biomedical Informatics}
\bvolume{41}(\bissue{5}),
\bfpage{706}--\blpage{716}
(\byear{2008}).
doi:\url{10.1016/j.jbi.2008.03.004}
\end{barticle}
\endbibitem

\bibitem{phpsparql}
\begin{botherref}
\oauthor{\bsnm{Gutteridge}, \binits{C.}}:
SPARQL RDF Library for PHP.
\url{http://graphite.ecs.soton.ac.uk/sparqllib/}
\end{botherref}
\endbibitem

\bibitem{Smith2007}
\begin{barticle}
\bauthor{\bsnm{Smith}, \binits{B.}},
\bauthor{\bsnm{Ashburner}, \binits{M.}},
\bauthor{\bsnm{Rosse}, \binits{C.}},
\bauthor{\bsnm{Bard}, \binits{J.}},
\bauthor{\bsnm{Bug}, \binits{W.}},
\bauthor{\bsnm{Ceusters}, \binits{W.}},
\bauthor{\bsnm{Goldberg}, \binits{L.J.}},
\bauthor{\bsnm{Eilbeck}, \binits{K.}},
\bauthor{\bsnm{Ireland}, \binits{A.}},
\bauthor{\bsnm{Mungall}, \binits{C.J.}},
\bauthor{\bsnm{Leontis}, \binits{N.}},
\bauthor{\bsnm{Serra}, \binits{P.R.}},
\bauthor{\bsnm{Ruttenberg}, \binits{A.}},
\bauthor{\bsnm{Sansone}, \binits{S.A.}},
\bauthor{\bsnm{Scheuermann}, \binits{R.H.}},
\bauthor{\bsnm{Shah}, \binits{N.}},
\bauthor{\bsnm{Whetzel}, \binits{P.L.}},
\bauthor{\bsnm{Lewis}, \binits{S.}}:
\batitle{The {OBO} {F}oundry: coordinated evolution of ontologies to support
  biomedical data integration}.
\bjtitle{Nat Biotech}
\bvolume{25}(\bissue{11}),
\bfpage{1251}--\blpage{1255}
(\byear{2007})
\end{barticle}
\endbibitem

\bibitem{sebastian2014}
\begin{botherref}
\oauthor{\bsnm{K\"{o}hler}, \binits{S.}},
\oauthor{\bsnm{Doelken}, \binits{S.C.}},
\oauthor{\bsnm{Ruef}, \binits{B.J.}},
\oauthor{\bsnm{Bauer}, \binits{S.}},
\oauthor{\bsnm{Washington}, \binits{N.}},
\oauthor{\bsnm{Westerfield}, \binits{M.}},
\oauthor{\bsnm{Gkoutos}, \binits{G.}},
\oauthor{\bsnm{Schofield}, \binits{P.}},
\oauthor{\bsnm{Smedley}, \binits{D.}},
\oauthor{\bsnm{Lewis}, \binits{S.E.}},
\oauthor{\bsnm{Robinson}, \binits{P.N.}},
\oauthor{\bsnm{Mungall}, \binits{C.J.}}:
Construction and accessibility of a cross-species phenotype ontology along with
  gene annotations for biomedical research.
F1000Research
\textbf{2}
(2013).
doi:\url{10.12688/f1000research.2-30.v1}
\end{botherref}
\endbibitem

\bibitem{Robinson2008}
\begin{barticle}
\bauthor{\bsnm{Robinson}, \binits{P.N.}},
\bauthor{\bsnm{Koehler}, \binits{S.}},
\bauthor{\bsnm{Bauer}, \binits{S.}},
\bauthor{\bsnm{Seelow}, \binits{D.}},
\bauthor{\bsnm{Horn}, \binits{D.}},
\bauthor{\bsnm{Mundlos}, \binits{S.}}:
\batitle{The human phenotype ontology: a tool for annotating and analyzing
  human hereditary disease.}
\bjtitle{American journal of human genetics}
\bvolume{83}(\bissue{5}),
\bfpage{610}--\blpage{615}
(\byear{2008}).
doi:\url{10.1016/j.ajhg.2008.09.017}
\end{barticle}
\endbibitem

\bibitem{Ashburner2000}
\begin{barticle}
\bauthor{\bsnm{Ashburner}, \binits{M.}},
\bauthor{\bsnm{Ball}, \binits{C.A.}},
\bauthor{\bsnm{Blake}, \binits{J.A.}},
\bauthor{\bsnm{Botstein}, \binits{D.}},
\bauthor{\bsnm{Butler}, \binits{H.}},
\bauthor{\bsnm{Cherry}, \binits{M.J.}},
\bauthor{\bsnm{Davis}, \binits{A.P.}},
\bauthor{\bsnm{Dolinski}, \binits{K.}},
\bauthor{\bsnm{Dwight}, \binits{S.S.}},
\bauthor{\bsnm{Eppig}, \binits{J.T.}},
\bauthor{\bsnm{Harris}, \binits{M.A.}},
\bauthor{\bsnm{Hill}, \binits{D.P.}},
\bauthor{\bsnm{Tarver}, \binits{L.I.}},
\bauthor{\bsnm{Kasarskis}, \binits{A.}},
\bauthor{\bsnm{Lewis}, \binits{S.}},
\bauthor{\bsnm{Matese}, \binits{J.C.}},
\bauthor{\bsnm{Richardson}, \binits{J.E.}},
\bauthor{\bsnm{Ringwald}, \binits{M.}},
\bauthor{\bsnm{Rubin}, \binits{G.M.}},
\bauthor{\bsnm{Sherlock}, \binits{G.}}:
\batitle{Gene ontology: tool for the unification of biology}.
\bjtitle{Nature Genetics}
\bvolume{25}(\bissue{1}),
\bfpage{25}--\blpage{29}
(\byear{2000}).
doi:\url{10.1038/75556}
\end{barticle}
\endbibitem

\bibitem{obo-id-policy}
\begin{botherref}
\oauthor{\bsnm{Ruttenberg}, \binits{A.}},
\oauthor{\bsnm{Courtot}, \binits{M.}},
\oauthor{\bsnm{Mungall}, \binits{C.J.}}:
OBO Foundry Identifier Policy.
\url{http://www.obofoundry.org/id-policy.shtml}
\end{botherref}
\endbibitem

\bibitem{Uniprot2007}
\begin{botherref}
\oauthor{\bsnm{{The Uniprot Consortium}}}:
The universal protein resource (uniprot).
Nucleic Acids Res
\textbf{35}(Database issue)
(2007)
\end{botherref}
\endbibitem

\bibitem{gwascentral}
\begin{barticle}
\bauthor{\bsnm{Beck}, \binits{T.}},
\bauthor{\bsnm{Hastings}, \binits{R.K.}},
\bauthor{\bsnm{Gollapudi}, \binits{S.}},
\bauthor{\bsnm{Free}, \binits{R.C.}},
\bauthor{\bsnm{Brookes}, \binits{A.J.}}:
\batitle{{GWAS} {C}entral: a comprehensive resource for the comparison and
  interrogation of genome-wide association studies}.
\bjtitle{Eur J Hum Genet}
\bvolume{22}(\bissue{7}),
\bfpage{949}--\blpage{52}
(\byear{2014})
\end{barticle}
\endbibitem

\bibitem{Mungall2010}
\begin{barticle}
\bauthor{\bsnm{Mungall}, \binits{C.}},
\bauthor{\bsnm{Gkoutos}, \binits{G.}},
\bauthor{\bsnm{Smith}, \binits{C.}},
\bauthor{\bsnm{Haendel}, \binits{M.}},
\bauthor{\bsnm{Lewis}, \binits{S.}},
\bauthor{\bsnm{Ashburner}, \binits{M.}}:
\batitle{Integrating phenotype ontologies across multiple species}.
\bjtitle{Genome Biology}
\bvolume{11}(\bissue{1}),
\bfpage{2}
(\byear{2010}).
doi:\url{10.1186/gb-2010-11-1-r2}
\end{barticle}
\endbibitem

\bibitem{Ontobee}
\begin{bchapter}
\bauthor{\bsnm{Z}, \binits{X.}},
\bauthor{\bsnm{C}, \binits{M.}},
\bauthor{\bsnm{A}, \binits{R.}},
\bauthor{\bsnm{Y}, \binits{H.}}:
\bctitle{Ontobee: A linked data server and browser for ontology terms}.
In: \bbtitle{Proceedings of International Conference on Biomedical Ontology},
pp. \bfpage{279}--\blpage{281}
(\byear{2011})
\end{bchapter}
\endbibitem

\bibitem{Tudose2013}
\begin{barticle}
\bauthor{\bsnm{Tudose}, \binits{I.}},
\bauthor{\bsnm{Hastings}, \binits{J.}},
\bauthor{\bsnm{Muthukrishnan}, \binits{V.}},
\bauthor{\bsnm{Owen}, \binits{G.}},
\bauthor{\bsnm{Turner}, \binits{S.}},
\bauthor{\bsnm{Dekker}, \binits{A.}},
\bauthor{\bsnm{Kale}, \binits{N.}},
\bauthor{\bsnm{Ennis}, \binits{M.}},
\bauthor{\bsnm{Steinbeck}, \binits{C.}}:
\batitle{Ontoquery: easy-to-use web-based owl querying}.
\bjtitle{Bioinformatics}
\bvolume{29}(\bissue{22}),
\bfpage{2955}--\blpage{2957}
(\byear{2013}).
doi:\url{10.1093/bioinformatics/btt514}
\end{barticle}
\endbibitem

\bibitem{Jupp2012}
\begin{barticle}
\bauthor{\bsnm{Jupp}, \binits{S.}},
\bauthor{\bsnm{Stevens}, \binits{R.}},
\bauthor{\bsnm{Hoehndorf}, \binits{R.}}:
\batitle{Logical gene ontology annotations (goal): exploring gene ontology
  annotations with owl}.
\bjtitle{Journal of Biomedical Semantics}
\bvolume{3}(\bissue{Suppl 1}),
\bfpage{3}
(\byear{2012}).
doi:\url{10.1186/2041-1480-3-S1-S3}
\end{barticle}
\endbibitem

\bibitem{Doms2005}
\begin{barticle}
\bauthor{\bsnm{Doms}, \binits{A.}},
\bauthor{\bsnm{Schroeder}, \binits{M.}}:
\batitle{{{G}o{P}ub{M}ed: exploring {P}ub{M}ed with the {G}ene {O}ntology}}.
\bjtitle{Nucleic Acids Res}
\bvolume{33}(\bissue{Web Server issue}),
\bfpage{783}--\blpage{786}
(\byear{2005})
\end{barticle}
\endbibitem

\bibitem{Nelson2004}
\begin{bchapter}
\bauthor{\bsnm{Nelson}, \binits{S.J.}},
\bauthor{\bsnm{Schopen}, \binits{M.}},
\bauthor{\bsnm{Savage}, \binits{A.G.}},
\bauthor{\bsnm{Schulman}, \binits{J.L.}},
\bauthor{\bsnm{Arluk}, \binits{N.}}:
\bctitle{The {MeSH} translation maintenance system: Structure, interface
  design, and implementation}.
In: \bbtitle{Proceedings of the 11th World Congress on Medical Informatics},
pp. \bfpage{67}--\blpage{69}.
\bpublisher{IOS Press},
\blocation{Amsterdam}
(\byear{2004})
\end{bchapter}
\endbibitem

\bibitem{Gaudan2008b}
\begin{botherref}
\oauthor{\bsnm{Gaudan}, \binits{S.}},
\oauthor{\bsnm{Jimeno~Yepes}, \binits{A.}},
\oauthor{\bsnm{Lee}, \binits{V.}},
\oauthor{\bsnm{Rebholz-Schuhmann}, \binits{D.}}:
Combining evidence, specificity, and proximity towards the normalization of
  gene ontology terms in text.
EURASIP journal on bioinformatics \& systems biology
(2008)
\end{botherref}
\endbibitem

\bibitem{Blaschke2005}
\begin{botherref}
\oauthor{\bsnm{Blaschke}, \binits{C.}},
\oauthor{\bsnm{Leon}, \binits{E.A.}},
\oauthor{\bsnm{Krallinger}, \binits{M.}},
\oauthor{\bsnm{Valencia}, \binits{A.}}:
Evaluation of {BioCreAtIvE} assessment of task 2.
BMC Bioinformatics
\textbf{6 Suppl 1}
(2005).
doi:\url{10.1186/1471-2105-6-S1-S16}
\end{botherref}
\endbibitem

\bibitem{sparqlentailment}
\begin{botherref}
\oauthor{\bsnm{Glimm}, \binits{B.}},
\oauthor{\bsnm{Ogbuji}, \binits{C.}}:
{SPARQL} 1.1 entailment regimes.
Recommendation,
World Wide Web Consortium (W3C)
(2013)
\end{botherref}
\endbibitem

\bibitem{Koehler2011}
\begin{barticle}
\bauthor{\bsnm{Kohler}, \binits{S.}},
\bauthor{\bsnm{Bauer}, \binits{S.}},
\bauthor{\bsnm{Mungall}, \binits{C.}},
\bauthor{\bsnm{Carletti}, \binits{G.}},
\bauthor{\bsnm{Smith}, \binits{C.}},
\bauthor{\bsnm{Schofield}, \binits{P.}},
\bauthor{\bsnm{Gkoutos}, \binits{G.}},
\bauthor{\bsnm{Robinson}, \binits{P.}}:
\batitle{Improving ontologies by automatic reasoning and evaluation of logical
  definitions}.
\bjtitle{BMC Bioinformatics}
\bvolume{12}(\bissue{1}),
\bfpage{418}
(\byear{2011}).
doi:\url{10.1186/1471-2105-12-418}
\end{barticle}
\endbibitem

\bibitem{Smith2005}
\begin{barticle}
\bauthor{\bsnm{Smith}, \binits{B.}},
\bauthor{\bsnm{Ceusters}, \binits{W.}},
\bauthor{\bsnm{Klagges}, \binits{B.}},
\bauthor{\bsnm{K\"{o}hler}, \binits{J.}},
\bauthor{\bsnm{Kumar}, \binits{A.}},
\bauthor{\bsnm{Lomax}, \binits{J.}},
\bauthor{\bsnm{Mungall}, \binits{C.}},
\bauthor{\bsnm{Neuhaus}, \binits{F.}},
\bauthor{\bsnm{Rector}, \binits{A.L.}},
\bauthor{\bsnm{Rosse}, \binits{C.}}:
\batitle{Relations in biomedical ontologies.}
\bjtitle{Genome Biol}
\bvolume{6}(\bissue{5}),
\bfpage{46}
(\byear{2005}).
doi:\url{10.1186/gb-2005-6-5-r46}
\end{barticle}
\endbibitem

\bibitem{Hoehndorf2013nbo}
\begin{barticle}
\bauthor{\bsnm{Hoehndorf}, \binits{R.}},
\bauthor{\bsnm{Hancock}, \binits{J.M.}},
\bauthor{\bsnm{Hardy}, \binits{N.W.}},
\bauthor{\bsnm{Mallon}, \binits{A.M.}},
\bauthor{\bsnm{Schofield}, \binits{P.N.}},
\bauthor{\bsnm{Gkoutos}, \binits{G.V.}}:
\batitle{Analyzing gene expression data in mice with the {N}euro {B}ehavior
  {O}ntology}.
\bjtitle{Mamm Genome}
\bvolume{25}(\bissue{1-2}),
\bfpage{32}--\blpage{40}
(\byear{2014})
\end{barticle}
\endbibitem

\end{thebibliography}

\newcommand{\BMCxmlcomment}[1]{}

\BMCxmlcomment{

<refgrp>

<bibl id="B1">
  <title><p>{OWL} 2: The next step for {OWL}</p></title>
  <aug>
    <au><snm>Grau</snm><fnm>B.</fnm></au>
    <au><snm>Horrocks</snm><fnm>I.</fnm></au>
    <au><snm>Motik</snm><fnm>B.</fnm></au>
    <au><snm>Parsia</snm><fnm>B.</fnm></au>
    <au><snm>Patelschneider</snm><fnm>P.</fnm></au>
    <au><snm>Sattler</snm><fnm>U.</fnm></au>
  </aug>
  <source>Web Semantics: Science, Services and Agents on the World Wide
  Web</source>
  <pubdate>2008</pubdate>
  <volume>6</volume>
  <issue>4</issue>
  <fpage>309</fpage>
  <lpage>-322</lpage>
  <url>http://dx.doi.org/10.1016/j.websem.2008.05.001</url>
</bibl>

<bibl id="B2">
  <title><p>{OBO} Flat File Format Syntax and Semantics and Mapping to {OWL}
  {W}eb {O}ntology {L}anguage</p></title>
  <aug>
    <au><snm>Horrocks</snm><fnm>I</fnm></au>
  </aug>
  <pubdate>2007</pubdate>
  <note>\url{http://www.cs.man.ac.uk/~horrocks/obo/}</note>
</bibl>

<bibl id="B3">
  <title><p>BioPortal: ontologies and integrated data resources at the click of
  a mouse.</p></title>
  <aug>
    <au><snm>Noy</snm><fnm>NF</fnm></au>
    <au><snm>Shah</snm><fnm>NH</fnm></au>
    <au><snm>Whetzel</snm><fnm>PL</fnm></au>
    <au><snm>Dai</snm><fnm>B</fnm></au>
    <au><snm>Dorf</snm><fnm>M</fnm></au>
    <au><snm>Griffith</snm><fnm>N</fnm></au>
    <au><snm>Jonquet</snm><fnm>C</fnm></au>
    <au><snm>Rubin</snm><fnm>DL</fnm></au>
    <au><snm>Storey</snm><fnm>MAA</fnm></au>
    <au><snm>Chute</snm><fnm>CG</fnm></au>
    <au><snm>Musen</snm><fnm>MA</fnm></au>
  </aug>
  <source>Nucleic acids research</source>
  <pubdate>2009</pubdate>
  <volume>37</volume>
  <issue>Web Server issue</issue>
  <fpage>W170</fpage>
  <lpage>-173</lpage>
  <url>http://dx.doi.org/10.1093/nar/gkp440</url>
</bibl>

<bibl id="B4">
  <title><p>The Ontology Lookup Service, a lightweight cross-platform tool for
  controlled vocabulary queries</p></title>
  <aug>
    <au><snm>Cote</snm><fnm>R</fnm></au>
    <au><snm>Jones</snm><fnm>P</fnm></au>
    <au><snm>Apweiler</snm><fnm>R</fnm></au>
    <au><snm>Hermjakob</snm><fnm>H</fnm></au>
  </aug>
  <source>BMC Bioinformatics</source>
  <pubdate>2006</pubdate>
  <volume>7</volume>
  <issue>1</issue>
  <fpage>97+</fpage>
  <url>http://dx.doi.org/10.1186/1471-2105-7-97</url>
</bibl>

<bibl id="B5">
  <title><p>The {EBI} {RDF} platform: linked open data for the life
  sciences</p></title>
  <aug>
    <au><snm>Jupp</snm><fnm>S</fnm></au>
    <au><snm>Malone</snm><fnm>J</fnm></au>
    <au><snm>Bolleman</snm><fnm>J</fnm></au>
    <au><snm>Brandizi</snm><fnm>M</fnm></au>
    <au><snm>Davies</snm><fnm>M</fnm></au>
    <au><snm>Garcia</snm><fnm>L</fnm></au>
    <au><snm>Gaulton</snm><fnm>A</fnm></au>
    <au><snm>Gehant</snm><fnm>S</fnm></au>
    <au><snm>Laibe</snm><fnm>C</fnm></au>
    <au><snm>Redaschi</snm><fnm>N</fnm></au>
    <au><snm>Wimalaratne</snm><fnm>SM</fnm></au>
    <au><snm>Martin</snm><fnm>M</fnm></au>
    <au><snm>Le Novère</snm><fnm>N</fnm></au>
    <au><snm>Parkinson</snm><fnm>H</fnm></au>
    <au><snm>Birney</snm><fnm>E</fnm></au>
    <au><snm>Jenkinson</snm><fnm>AM</fnm></au>
  </aug>
  <source>Bioinformatics</source>
  <pubdate>2014</pubdate>
  <volume>30</volume>
  <issue>9</issue>
  <fpage>1338</fpage>
  <lpage>1339</lpage>
  <url>http://bioinformatics.oxfordjournals.org/content/30/9/1338.abstract</url>
</bibl>

<bibl id="B6">
  <title><p>Ontology-based data access: a study through disjunctive datalog,
  CSP, and MMSNP</p></title>
  <aug>
    <au><snm>Bienvenu</snm><fnm>M</fnm></au>
    <au><snm>Cate</snm><fnm>B</fnm></au>
    <au><snm>Lutz</snm><fnm>C</fnm></au>
    <au><snm>Wolter</snm><fnm>F</fnm></au>
  </aug>
  <source>PODS</source>
  <pubdate>2013</pubdate>
  <fpage>213</fpage>
  <lpage>224</lpage>
</bibl>

<bibl id="B7">
  <title><p>The Combined Approach to Ontology-Based Data Access</p></title>
  <aug>
    <au><snm>Kontchakov</snm><fnm>R</fnm></au>
    <au><snm>Lutz</snm><fnm>C</fnm></au>
    <au><snm>Toman</snm><fnm>D</fnm></au>
    <au><snm>Wolter</snm><fnm>F</fnm></au>
    <au><snm>Zakharyaschev</snm><fnm>M</fnm></au>
  </aug>
  <source>IJCAI</source>
  <pubdate>2011</pubdate>
  <fpage>2656</fpage>
  <lpage>2661</lpage>
</bibl>

<bibl id="B8">
  <title><p>Interoperability between biomedical ontologies through relation
  expansion, upper-level ontologies and automatic reasoning</p></title>
  <aug>
    <au><snm>Hoehndorf</snm><fnm>R</fnm></au>
    <au><snm>Dumontier</snm><fnm>M</fnm></au>
    <au><snm>Oellrich</snm><fnm>A</fnm></au>
    <au><snm>Rebholz Schuhmann</snm><fnm>D</fnm></au>
    <au><snm>Schofield</snm><fnm>PN</fnm></au>
    <au><snm>Gkoutos</snm><fnm>GV</fnm></au>
  </aug>
  <source>PLOS ONE</source>
  <pubdate>2011</pubdate>
  <volume>6</volume>
  <issue>7</issue>
  <fpage>e22006</fpage>
</bibl>

<bibl id="B9">
  <title><p>PhenomeNET: a whole-phenome approach to disease gene
  discovery</p></title>
  <aug>
    <au><snm>Hoehndorf</snm><fnm>R</fnm></au>
    <au><snm>Schofield</snm><fnm>PN</fnm></au>
    <au><snm>Gkoutos</snm><fnm>GV</fnm></au>
  </aug>
  <source>Nucleic Acids Research</source>
  <pubdate>2011</pubdate>
  <volume>39</volume>
  <issue>18</issue>
  <fpage>e119</fpage>
  <url>http://nar.oxfordjournals.org/content/39/18/e119</url>
</bibl>

<bibl id="B10">
  <title><p>The Ontological Level</p></title>
  <aug>
    <au><snm>Guarino</snm><fnm>N</fnm></au>
  </aug>
  <source>Philosophy and the Cognitive Sciences</source>
  <publisher>Vienna: H\"{o}lder-Pichler-Tempsky</publisher>
  <editor>Casati, R. and Smith, B. and White, G.</editor>
  <pubdate>1994</pubdate>
  <fpage>443</fpage>
  <lpage>-456</lpage>
</bibl>

<bibl id="B11">
  <title><p>Igniting the {OWL} 1.1 Touch Paper: The {OWL} {API}</p></title>
  <aug>
    <au><snm>Horridge</snm><fnm>M</fnm></au>
    <au><snm>Bechhofer</snm><fnm>S</fnm></au>
    <au><snm>Noppens</snm><fnm>O</fnm></au>
  </aug>
  <source>Proceedings of OWLED 2007: Third International Workshop on OWL
  Experiences and Directions</source>
  <pubdate>2007</pubdate>
</bibl>

<bibl id="B12">
  <title><p>Unchain My $\mathcal{EL}$ Reasoner</p></title>
  <aug>
    <au><snm>Kazakov</snm><fnm>Y</fnm></au>
    <au><snm>Kr{\"o}tzsch</snm><fnm>M</fnm></au>
    <au><snm>Siman\v{c}\'{i}k</snm><fnm>F</fnm></au>
  </aug>
  <source>Proceedings of the 23rd International Workshop on Description Logics
  (DL'10)</source>
  <publisher>CEUR-WS.org</publisher>
  <series><title><p>CEUR Workshop Proceedings</p></title></series>
  <pubdate>2011</pubdate>
</bibl>

<bibl id="B13">
  <title><p>OWL 2 Web Ontology Language: Profiles</p></title>
  <aug>
    <au><snm>Motik</snm><fnm>B</fnm></au>
    <au><snm>Grau</snm><fnm>BC</fnm></au>
    <au><snm>Horrocks</snm><fnm>I</fnm></au>
    <au><snm>Wu</snm><fnm>Z</fnm></au>
    <au><snm>Fokoue</snm><fnm>A</fnm></au>
    <au><snm>Lutz</snm><fnm>C</fnm></au>
  </aug>
  <source>Recommendation</source>
  <pubdate>2009</pubdate>
</bibl>

<bibl id="B14">
  <title><p>A common layer of interoperability for biomedical ontologies based
  on {OWL} {EL}</p></title>
  <aug>
    <au><snm>Hoehndorf</snm><fnm>R</fnm></au>
    <au><snm>Dumontier</snm><fnm>M</fnm></au>
    <au><snm>Oellrich</snm><fnm>A</fnm></au>
    <au><snm>Wimalaratne</snm><fnm>S</fnm></au>
    <au><snm>Rebholz Schuhmann</snm><fnm>D</fnm></au>
    <au><snm>Schofield</snm><fnm>P</fnm></au>
    <au><snm>Gkoutos</snm><fnm>GV</fnm></au>
  </aug>
  <source>Bioinformatics</source>
  <pubdate>2011</pubdate>
  <volume>27</volume>
  <issue>7</issue>
  <fpage>1001</fpage>
  <lpage>-1008</lpage>
</bibl>

<bibl id="B15">
  <title><p>{Hypertableau Reasoning for Description Logics}</p></title>
  <aug>
    <au><snm>Motik</snm><fnm>B</fnm></au>
    <au><snm>Shearer</snm><fnm>R</fnm></au>
    <au><snm>Horrocks</snm><fnm>I</fnm></au>
  </aug>
  <source>Journal of Artificial Intelligence Research</source>
  <pubdate>2009</pubdate>
  <volume>36</volume>
  <fpage>165</fpage>
  <lpage>-228</lpage>
</bibl>

<bibl id="B16">
  <title><p>Pellet: An {OWL} {DL} Reasoner</p></title>
  <aug>
    <au><snm>Sirin</snm><fnm>E</fnm></au>
    <au><snm>Parsia</snm><fnm>B</fnm></au>
  </aug>
  <source>Proceedings of the 2004 International Workshop on Description Logics,
  DL2004, Whistler, British Columbia, Canada, Jun 6-8</source>
  <publisher>Aachen, Germany: CEUR-WS.org</publisher>
  <editor>Haarslev, Volker and M{\"{o}}ller, Ralf</editor>
  <series><title><p>CEUR Workshop Proceedings</p></title></series>
  <pubdate>2004</pubdate>
  <volume>104</volume>
</bibl>

<bibl id="B17">
  <title><p>{The Manchester OWL Syntax}</p></title>
  <aug>
    <au><snm>Horridge</snm><fnm>M.</fnm></au>
    <au><snm>Drummond</snm><fnm>N.</fnm></au>
    <au><snm>Goodwin</snm><fnm>J.</fnm></au>
    <au><snm>Rector</snm><fnm>A.</fnm></au>
    <au><snm>Stevens</snm><fnm>R.</fnm></au>
    <au><snm>Wang</snm><fnm>H.H.</fnm></au>
  </aug>
  <source>Proc. of the 2006 OWL Experiences and Directions Workshop
  (OWL-ED2006)</source>
  <pubdate>2006</pubdate>
</bibl>

<bibl id="B18">
  <title><p>{The JavaScript Object Notation (JSON) Data Interchange
  Format}</p></title>
  <aug>
    <au><snm>Bray</snm><fnm>T.</fnm></au>
  </aug>
  <source>RFC 7159 (Proposed Standard)</source>
  <publisher>IETF</publisher>
  <series><title><p>Request for Comments</p></title></series>
  <pubdate>2014</pubdate>
  <issue>7159</issue>
  <url>http://www.ietf.org/rfc/rfc7159.txt</url>
</bibl>

<bibl id="B19">
  <title><p>{jQuery}: The write less, do more, JavaScript library</p></title>
  <aug>
    <au><cnm>{The jQuery Project}</cnm></au>
  </aug>
  <source>\url{http://jquery.com}</source>
</bibl>

<bibl id="B20">
  <title><p>Ajax: A New Approach to Web Applications</p></title>
  <aug>
    <au><snm>Garrett</snm><fnm>JJ</fnm></au>
  </aug>
  <source>\url{http://www.adaptivepath.com/ideas/essays/archives/000385.php}</source>
  <pubdate>2005</pubdate>
</bibl>

<bibl id="B21">
  <title><p>Apache Lucene</p></title>
  <aug>
    <au><cnm>{The Apache Software Foundation}</cnm></au>
  </aug>
  <source>\url{http://lucene.apache.org}</source>
</bibl>

<bibl id="B22">
  <title><p>2014 MEDLINE/PubMed Baseline Distribution</p></title>
  <aug>
    <au><cnm>{U.S. National Library of Medicine}</cnm></au>
  </aug>
  <source>\url{http://www.nlm.nih.gov/bsd/licensee/2014_stats/baseline_doc.html}</source>
</bibl>

<bibl id="B23">
  <title><p>PubMed Central</p></title>
  <aug>
    <au><cnm>{U.S. National Library of Medicine}</cnm></au>
  </aug>
  <source>\url{http://www.ncbi.nlm.nih.gov/pmc/}</source>
</bibl>

<bibl id="B24">
  <title><p>{Bio2RDF}: Towards a mashup to build bioinformatics knowledge
  systems</p></title>
  <aug>
    <au><snm>Belleau</snm><fnm>F.</fnm></au>
    <au><snm>Nolin</snm><fnm>M.</fnm></au>
    <au><snm>Tourigny</snm><fnm>N.</fnm></au>
    <au><snm>Rigault</snm><fnm>P.</fnm></au>
    <au><snm>Morissette</snm><fnm>J.</fnm></au>
  </aug>
  <source>Journal of Biomedical Informatics</source>
  <publisher>Centre de Recherche du CHUL, Universit\'{e} Laval, 2705 Boulevard
  Laurier, Que., Canada G1V 4G2.</publisher>
  <pubdate>2008</pubdate>
  <volume>41</volume>
  <issue>5</issue>
  <fpage>706</fpage>
  <lpage>-716</lpage>
  <url>http://dx.doi.org/10.1016/j.jbi.2008.03.004</url>
</bibl>

<bibl id="B25">
  <title><p>SPARQL RDF Library for PHP</p></title>
  <aug>
    <au><snm>Gutteridge</snm><fnm>C</fnm></au>
  </aug>
  <source>\url{http://graphite.ecs.soton.ac.uk/sparqllib/}</source>
</bibl>

<bibl id="B26">
  <title><p>The {OBO} {F}oundry: coordinated evolution of ontologies to support
  biomedical data integration</p></title>
  <aug>
    <au><snm>Smith</snm><fnm>B</fnm></au>
    <au><snm>Ashburner</snm><fnm>M</fnm></au>
    <au><snm>Rosse</snm><fnm>C</fnm></au>
    <au><snm>Bard</snm><fnm>J</fnm></au>
    <au><snm>Bug</snm><fnm>W</fnm></au>
    <au><snm>Ceusters</snm><fnm>W</fnm></au>
    <au><snm>Goldberg</snm><fnm>LJ</fnm></au>
    <au><snm>Eilbeck</snm><fnm>K</fnm></au>
    <au><snm>Ireland</snm><fnm>A</fnm></au>
    <au><snm>Mungall</snm><fnm>CJ</fnm></au>
    <au><snm>Leontis</snm><fnm>N</fnm></au>
    <au><snm>Serra</snm><fnm>PR</fnm></au>
    <au><snm>Ruttenberg</snm><fnm>A</fnm></au>
    <au><snm>Sansone</snm><fnm>SA</fnm></au>
    <au><snm>Scheuermann</snm><fnm>RH</fnm></au>
    <au><snm>Shah</snm><fnm>N</fnm></au>
    <au><snm>Whetzel</snm><fnm>PL</fnm></au>
    <au><snm>Lewis</snm><fnm>S</fnm></au>
  </aug>
  <source>Nat Biotech</source>
  <publisher>Nature Publishing Group</publisher>
  <pubdate>2007</pubdate>
  <volume>25</volume>
  <issue>11</issue>
  <fpage>1251</fpage>
  <lpage>-1255</lpage>
</bibl>

<bibl id="B27">
  <title><p>Construction and accessibility of a cross-species phenotype
  ontology along with gene annotations for biomedical research.</p></title>
  <aug>
    <au><snm>K\"{o}hler</snm><fnm>S</fnm></au>
    <au><snm>Doelken</snm><fnm>SC</fnm></au>
    <au><snm>Ruef</snm><fnm>BJ</fnm></au>
    <au><snm>Bauer</snm><fnm>S</fnm></au>
    <au><snm>Washington</snm><fnm>N</fnm></au>
    <au><snm>Westerfield</snm><fnm>M</fnm></au>
    <au><snm>Gkoutos</snm><fnm>G</fnm></au>
    <au><snm>Schofield</snm><fnm>P</fnm></au>
    <au><snm>Smedley</snm><fnm>D</fnm></au>
    <au><snm>Lewis</snm><fnm>SE</fnm></au>
    <au><snm>Robinson</snm><fnm>PN</fnm></au>
    <au><snm>Mungall</snm><fnm>CJ</fnm></au>
  </aug>
  <source>F1000Research</source>
  <pubdate>2013</pubdate>
  <volume>2</volume>
  <url>http://dx.doi.org/10.12688/f1000research.2-30.v1</url>
</bibl>

<bibl id="B28">
  <title><p>The Human Phenotype Ontology: a tool for annotating and analyzing
  human hereditary disease.</p></title>
  <aug>
    <au><snm>Robinson</snm><fnm>P. N.</fnm></au>
    <au><snm>Koehler</snm><fnm>S.</fnm></au>
    <au><snm>Bauer</snm><fnm>S.</fnm></au>
    <au><snm>Seelow</snm><fnm>D.</fnm></au>
    <au><snm>Horn</snm><fnm>D.</fnm></au>
    <au><snm>Mundlos</snm><fnm>S.</fnm></au>
  </aug>
  <source>American journal of human genetics</source>
  <pubdate>2008</pubdate>
  <volume>83</volume>
  <issue>5</issue>
  <fpage>610</fpage>
  <lpage>-615</lpage>
  <url>http://dx.doi.org/10.1016/j.ajhg.2008.09.017</url>
</bibl>

<bibl id="B29">
  <title><p>Gene ontology: tool for the unification of biology</p></title>
  <aug>
    <au><snm>Ashburner</snm><fnm>M</fnm></au>
    <au><snm>Ball</snm><fnm>CA</fnm></au>
    <au><snm>Blake</snm><fnm>JA</fnm></au>
    <au><snm>Botstein</snm><fnm>D</fnm></au>
    <au><snm>Butler</snm><fnm>H</fnm></au>
    <au><snm>Cherry</snm><fnm>MJ</fnm></au>
    <au><snm>Davis</snm><fnm>AP</fnm></au>
    <au><snm>Dolinski</snm><fnm>K</fnm></au>
    <au><snm>Dwight</snm><fnm>SS</fnm></au>
    <au><snm>Eppig</snm><fnm>JT</fnm></au>
    <au><snm>Harris</snm><fnm>MA</fnm></au>
    <au><snm>Hill</snm><fnm>DP</fnm></au>
    <au><snm>Tarver</snm><fnm>LI</fnm></au>
    <au><snm>Kasarskis</snm><fnm>A</fnm></au>
    <au><snm>Lewis</snm><fnm>S</fnm></au>
    <au><snm>Matese</snm><fnm>JC</fnm></au>
    <au><snm>Richardson</snm><fnm>JE</fnm></au>
    <au><snm>Ringwald</snm><fnm>M</fnm></au>
    <au><snm>Rubin</snm><fnm>GM</fnm></au>
    <au><snm>Sherlock</snm><fnm>G</fnm></au>
  </aug>
  <source>Nature Genetics</source>
  <pubdate>2000</pubdate>
  <volume>25</volume>
  <issue>1</issue>
  <fpage>25</fpage>
  <lpage>-29</lpage>
  <url>http://dx.doi.org/10.1038/75556</url>
</bibl>

<bibl id="B30">
  <title><p>OBO Foundry Identifier Policy</p></title>
  <aug>
    <au><snm>Ruttenberg</snm><fnm>A</fnm></au>
    <au><snm>Courtot</snm><fnm>M</fnm></au>
    <au><snm>Mungall</snm><fnm>CJ</fnm></au>
  </aug>
  <source>\url{http://www.obofoundry.org/id-policy.shtml}</source>
</bibl>

<bibl id="B31">
  <title><p>The Universal Protein Resource (UniProt).</p></title>
  <aug>
    <au><cnm>{The Uniprot Consortium}</cnm></au>
  </aug>
  <source>Nucleic Acids Res</source>
  <pubdate>2007</pubdate>
  <volume>35</volume>
  <issue>Database issue</issue>
  <url>http://view.ncbi.nlm.nih.gov/pubmed/17142230</url>
</bibl>

<bibl id="B32">
  <title><p>{GWAS} {C}entral: a comprehensive resource for the comparison and
  interrogation of genome-wide association studies</p></title>
  <aug>
    <au><snm>Beck</snm><fnm>T.</fnm></au>
    <au><snm>Hastings</snm><fnm>R. K.</fnm></au>
    <au><snm>Gollapudi</snm><fnm>S.</fnm></au>
    <au><snm>Free</snm><fnm>R. C.</fnm></au>
    <au><snm>Brookes</snm><fnm>A. J.</fnm></au>
  </aug>
  <source>Eur J Hum Genet</source>
  <pubdate>2014</pubdate>
  <volume>22</volume>
  <issue>7</issue>
  <fpage>949</fpage>
  <lpage>52</lpage>
</bibl>

<bibl id="B33">
  <title><p>Integrating phenotype ontologies across multiple
  species</p></title>
  <aug>
    <au><snm>Mungall</snm><fnm>C</fnm></au>
    <au><snm>Gkoutos</snm><fnm>G</fnm></au>
    <au><snm>Smith</snm><fnm>C</fnm></au>
    <au><snm>Haendel</snm><fnm>M</fnm></au>
    <au><snm>Lewis</snm><fnm>S</fnm></au>
    <au><snm>Ashburner</snm><fnm>M</fnm></au>
  </aug>
  <source>Genome Biology</source>
  <pubdate>2010</pubdate>
  <volume>11</volume>
  <issue>1</issue>
  <fpage>R2+</fpage>
  <url>http://dx.doi.org/10.1186/gb-2010-11-1-r2</url>
</bibl>

<bibl id="B34">
  <title><p>Ontobee: A Linked Data Server and Browser for Ontology
  Terms</p></title>
  <aug>
    <au><snm>Z</snm><fnm>X</fnm></au>
    <au><snm>C</snm><fnm>M</fnm></au>
    <au><snm>A</snm><fnm>R</fnm></au>
    <au><snm>Y</snm><fnm>H</fnm></au>
  </aug>
  <source>Proceedings of International Conference on Biomedical
  Ontology</source>
  <pubdate>2011</pubdate>
  <fpage>279</fpage>
  <lpage>281</lpage>
</bibl>

<bibl id="B35">
  <title><p>OntoQuery: easy-to-use web-based OWL querying</p></title>
  <aug>
    <au><snm>Tudose</snm><fnm>I</fnm></au>
    <au><snm>Hastings</snm><fnm>J</fnm></au>
    <au><snm>Muthukrishnan</snm><fnm>V</fnm></au>
    <au><snm>Owen</snm><fnm>G</fnm></au>
    <au><snm>Turner</snm><fnm>S</fnm></au>
    <au><snm>Dekker</snm><fnm>A</fnm></au>
    <au><snm>Kale</snm><fnm>N</fnm></au>
    <au><snm>Ennis</snm><fnm>M</fnm></au>
    <au><snm>Steinbeck</snm><fnm>C</fnm></au>
  </aug>
  <source>Bioinformatics</source>
  <pubdate>2013</pubdate>
  <volume>29</volume>
  <issue>22</issue>
  <fpage>2955</fpage>
  <lpage>2957</lpage>
  <url>http://bioinformatics.oxfordjournals.org/content/29/22/2955.abstract</url>
</bibl>

<bibl id="B36">
  <title><p>Logical Gene Ontology Annotations (GOAL): exploring gene ontology
  annotations with OWL</p></title>
  <aug>
    <au><snm>Jupp</snm><fnm>S</fnm></au>
    <au><snm>Stevens</snm><fnm>R</fnm></au>
    <au><snm>Hoehndorf</snm><fnm>R</fnm></au>
  </aug>
  <source>Journal of Biomedical Semantics</source>
  <pubdate>2012</pubdate>
  <volume>3</volume>
  <issue>Suppl 1</issue>
  <fpage>S3</fpage>
  <url>http://www.jbiomedsem.com/supplements/3/S1/S3</url>
</bibl>

<bibl id="B37">
  <title><p>{{G}o{P}ub{M}ed: exploring {P}ub{M}ed with the {G}ene
  {O}ntology}</p></title>
  <aug>
    <au><snm>Doms</snm><fnm>A.</fnm></au>
    <au><snm>Schroeder</snm><fnm>M.</fnm></au>
  </aug>
  <source>Nucleic Acids Res</source>
  <pubdate>2005</pubdate>
  <volume>33</volume>
  <issue>Web Server issue</issue>
  <fpage>783</fpage>
  <lpage>-786</lpage>
</bibl>

<bibl id="B38">
  <title><p>The {MeSH} translation maintenance system: Structure, interface
  design, and implementation</p></title>
  <aug>
    <au><snm>Nelson</snm><fnm>S. J.</fnm></au>
    <au><snm>Schopen</snm><fnm>M.</fnm></au>
    <au><snm>Savage</snm><fnm>A. G.</fnm></au>
    <au><snm>Schulman</snm><fnm>J. L.</fnm></au>
    <au><snm>Arluk</snm><fnm>N.</fnm></au>
  </aug>
  <source>Proceedings of the 11th World Congress on Medical
  Informatics</source>
  <publisher>Amsterdam: IOS Press</publisher>
  <pubdate>2004</pubdate>
  <fpage>67</fpage>
  <lpage>-69</lpage>
</bibl>

<bibl id="B39">
  <title><p>Combining Evidence, Specificity, and Proximity towards the
  Normalization of Gene Ontology Terms in Text.</p></title>
  <aug>
    <au><snm>Gaudan</snm><fnm>S.</fnm></au>
    <au><snm>Jimeno Yepes</snm><fnm>A.</fnm></au>
    <au><snm>Lee</snm><fnm>V.</fnm></au>
    <au><snm>Rebholz Schuhmann</snm><fnm>D.</fnm></au>
  </aug>
  <source>EURASIP journal on bioinformatics \& systems biology</source>
  <publisher>European Bioinformatics Institute, Cambridge CB10 1SD,
  UK.</publisher>
  <pubdate>2008</pubdate>
  <url>http://view.ncbi.nlm.nih.gov/pubmed/18437221</url>
</bibl>

<bibl id="B40">
  <title><p>Evaluation of {BioCreAtIvE} assessment of task 2.</p></title>
  <aug>
    <au><snm>Blaschke</snm><fnm>C.</fnm></au>
    <au><snm>Leon</snm><fnm>E. A.</fnm></au>
    <au><snm>Krallinger</snm><fnm>M.</fnm></au>
    <au><snm>Valencia</snm><fnm>A.</fnm></au>
  </aug>
  <source>BMC Bioinformatics</source>
  <pubdate>2005</pubdate>
  <volume>6 Suppl 1</volume>
  <url>http://dx.doi.org/10.1186/1471-2105-6-S1-S16</url>
</bibl>

<bibl id="B41">
  <title><p>{SPARQL} 1.1 Entailment Regimes</p></title>
  <aug>
    <au><snm>Glimm</snm><fnm>B</fnm></au>
    <au><snm>Ogbuji</snm><fnm>C</fnm></au>
  </aug>
  <source>Recommendation</source>
  <pubdate>2013</pubdate>
</bibl>

<bibl id="B42">
  <title><p>Improving ontologies by automatic reasoning and evaluation of
  logical definitions</p></title>
  <aug>
    <au><snm>Kohler</snm><fnm>S</fnm></au>
    <au><snm>Bauer</snm><fnm>S</fnm></au>
    <au><snm>Mungall</snm><fnm>C</fnm></au>
    <au><snm>Carletti</snm><fnm>G</fnm></au>
    <au><snm>Smith</snm><fnm>C</fnm></au>
    <au><snm>Schofield</snm><fnm>P</fnm></au>
    <au><snm>Gkoutos</snm><fnm>G</fnm></au>
    <au><snm>Robinson</snm><fnm>P</fnm></au>
  </aug>
  <source>BMC Bioinformatics</source>
  <pubdate>2011</pubdate>
  <volume>12</volume>
  <issue>1</issue>
  <fpage>418</fpage>
  <url>http://www.biomedcentral.com/1471-2105/12/418</url>
</bibl>

<bibl id="B43">
  <title><p>Relations in biomedical ontologies.</p></title>
  <aug>
    <au><snm>Smith</snm><fnm>B.</fnm></au>
    <au><snm>Ceusters</snm><fnm>W.</fnm></au>
    <au><snm>Klagges</snm><fnm>B.</fnm></au>
    <au><snm>K\"{o}hler</snm><fnm>J.</fnm></au>
    <au><snm>Kumar</snm><fnm>A.</fnm></au>
    <au><snm>Lomax</snm><fnm>J.</fnm></au>
    <au><snm>Mungall</snm><fnm>C.</fnm></au>
    <au><snm>Neuhaus</snm><fnm>F.</fnm></au>
    <au><snm>Rector</snm><fnm>A. L.</fnm></au>
    <au><snm>Rosse</snm><fnm>C.</fnm></au>
  </aug>
  <source>Genome Biol</source>
  <pubdate>2005</pubdate>
  <volume>6</volume>
  <issue>5</issue>
  <fpage>R46</fpage>
  <url>http://dx.doi.org/10.1186/gb-2005-6-5-r46</url>
</bibl>

<bibl id="B44">
  <title><p>Analyzing gene expression data in mice with the {N}euro {B}ehavior
  {O}ntology</p></title>
  <aug>
    <au><snm>Hoehndorf</snm><fnm>R.</fnm></au>
    <au><snm>Hancock</snm><fnm>J. M.</fnm></au>
    <au><snm>Hardy</snm><fnm>N. W.</fnm></au>
    <au><snm>Mallon</snm><fnm>A. M.</fnm></au>
    <au><snm>Schofield</snm><fnm>P. N.</fnm></au>
    <au><snm>Gkoutos</snm><fnm>G. V.</fnm></au>
  </aug>
  <source>Mamm Genome</source>
  <pubdate>2014</pubdate>
  <volume>25</volume>
  <issue>1-2</issue>
  <fpage>32</fpage>
  <lpage>40</lpage>
</bibl>

</refgrp>
} 

\end{document}